\documentclass[12pt,epsf]{article}
\usepackage{epsf}

\newcommand{\dlt}{\delta}

\newcommand{\gm}{\gamma}

\newcommand{\tht}{\theta}

\newcommand{\lmd}{\lambda}
\newcommand{\Lmd}{\Lambda}
\newcommand{\sgm}{\sigma}
\newcommand{\vph}{\varphi}

\newcommand{\be}{\begin{equation}}
\newcommand{\ee}{\end{equation}}
\newcommand{\bea}{\begin{eqnarray}}
\newcommand{\eea}{\end{eqnarray}}
\newcommand{\eql}{\!\!\!&=\!\!\!&}

\newcommand{\defa}{\!\!\!&\equiv\!\!\!&}

\newcommand{\mtrx}[4]{\brkt{\begin{array}{cc}#1&#2\\#3&#4\end{array}}}

\newcommand{\exch}{\leftrightarrow}
\newcommand{\simgt}{\stackrel{>}{{}_\sim}}

\newcommand{\tl}[1]{\tilde{#1}}

\newcommand{\tr}{{\rm tr}}

\newcommand{\hc}{{\rm h.c.}}
\newcommand{\ie}{{\it i.e.}}

\newcommand{\mrl}{\tl{m}_{\rm RL}}
\newcommand{\mrr}{\tl{m}_{\rm RR}}
\newcommand{\mll}{\tl{m}_{\rm LL}}
\newcommand{\Lsb}{\Lambda_{\rm S}}
\newcommand{\Luv}{\Lambda_{\rm UV}}
\newcommand{\Mp}{M_{\rm P}}

\newcommand{\vev}[1]{\langle #1 \rangle}

\newcommand{\brkt}[1]{\left( #1 \right)}
\newcommand{\brc}[1]{\left\{ #1 \right\}}
\newcommand{\sbk}[1]{\left[ #1 \right]}
\newcommand{\abs}[1]{\left| #1 \right|}

\renewcommand{\Im}{{\rm Im}}

\newcommand{\cL}{{\cal L}}
\newcommand{\cM}{{\cal M}}

\newcommand{\cO}{{\cal O}}

\newcommand{\cW}{{\cal W}}

\newcommand{\NP}[1]{{\it Nucl.~Phys.}~{\bf #1}}
\newcommand{\PL}[1]{{\it Phys.~Lett.}~{\bf #1}}

\newcommand{\PR}[1]{{\it Phys.~Rev.}~{\bf #1}}
\newcommand{\PRL}[1]{{\it Phys.~Rev.~Lett.}~{\bf #1}}

\newcommand{\JH}[1]{{\it JHEP}~{\bf #1}}

\begin{document}

\begin{titlepage}
\null
\begin{flushright}
 {\tt hep-ph/0308115}\\
TU-696
\\
August, 2003
\end{flushright}

\vskip 2cm
\begin{center}
\baselineskip 0.8cm
{\LARGE \bf CP violation in models with TeV-scale SUSY breaking}

\lineskip .75em
\vskip 2.5cm

\normalsize

{\large\bf Motoi Endo}{
\def\thefootnote{\fnsymbol{footnote}
}\footnote[1]{\it  e-mail address: 
endo@tuhep.phys.tohoku.ac.jp}}
~and~~
{\large\bf Yutaka Sakamura}{
\def\thefootnote{\fnsymbol{footnote}
}\footnote[2]{\it  e-mail address:
sakamura@tuhep.phys.tohoku.ac.jp}}

\vskip 1.5em

{\it Department of Physics, Tohoku University\\ 
Sendai 980-8578, Japan}

\vspace{18mm}

{\bf Abstract}\\[5mm]
{\parbox{13cm}{\hspace{5mm} \small
We estimate contributions of the goldstino supermultiplet 
to the electric dipole moment (EDM) 
in the case that the SUSY breaking scale~$\Lsb$ is of order the TeV scale. 
We found that such contributions can saturate the experimental bound 
if $\Lsb$ is close to the soft mass scale. 
We also discuss EDM in the gaugino mediated scenario on the warped geometry 
as an example of models with the TeV-scale SUSY breaking. 
}}

\end{center}

\end{titlepage}

\clearpage

\section{Introduction}
Supersymmetry (SUSY) is one of the most promising candidate for the physics 
beyond the standard model. 
It must be, however, broken at low-energy since no superparticles have 
been observed yet. 
Although the SUSY breaking scale~$\Lsb$ is considered to be much higher 
than the weak scale~$G_{\rm F}^{-1/2}$ in the conventional scenarios, 
an experimental lower bound on it is relatively weak, \ie, 
$\Lsb\simgt G_{\rm F}^{-1/2}$, which come from 
the collider experiments\footnote{Constraints from cosmology and astrophysics 
are somewhat weaker \cite{brignole2,gherghetta}. 
} \cite{brignole1}. 
In this paper, we will consider a possibility that $\Lsb$ is close to 
its lower bound, typically the TeV scale. 
In such a case, we should take into account not only the fields of 
the minimal supersymmetric standard model (MSSM), but also 
the {\it goldstino} supermultiplet 
as the physical degrees of freedom in the low-energy effective theory. 
Couplings between the goldstino supermultiplet and the MSSM fields 
are generally suppressed by negative powers of $\Lsb$ \cite{brignole2}, 
and thus they become relevant to the low-energy phenomena 
\cite{brignole3,brignole4,brignole5} 
when $\Lsb$ is close to the soft SUSY-breaking mass scale~$\tl{m}$. 

Among low-energy phenomena, CP violation is very sensitive to the physics 
beyond the weak scale. 
Due to the appearance of the goldstino supermultiplet, 
there are additional sources of CP violation besides the MSSM ones. 
Such new CP violating sources generally induce sizable contributions 
to the electric dipole moments (EDMs) and thus in this paper, we will 
estimate the contributions of the goldstino supermultiplet to the EDMs. 

The paper is organized as follows. 
In the next section, we will provide notations and an effective theory 
used in our discussion. 
In Sect.~\ref{est_EDM}, we will calculate the contributions of the goldstino 
supermultiplet to the electron EDM. 
In Sect.~\ref{warped_GM}, we will estimate EDM in the gaugino mediation model  
with the warped geometry which is an example of models with the TeV-scale 
SUSY breaking. 
Sect.~\ref{conclusion} is devoted to the summary.

\section{Effective theory}
We will basically follow the effective theory approach 
in Ref.\cite{brignole3}. 
In general, the goldstino is absorbed into the gravitino and becomes 
a longitudinal component of the massive gravitino. 
In the case considered here, the gravitino mass~$m_{3/2}$ is several orders of 
magnitude below the eV scale. 
Therefore, a typical energy scale is much higher than $m_{3/2}$ 
in most of the physical processes, 
and the gravitino interactions can be well approximated 
by those of the goldstino in the global SUSY limit, 
thanks to the supersymmetric equivalence theorem \cite{equivTHM}. 
So we will work in the global SUSY limit in the following. 

The relevant fields to our discussion consist of 
the following supermultiplets\footnote{
We use the notations of Ref.\cite{WB} in this paper. 
}. 
The left-handed lepton doublet~$L=(\tl{l},l,F^l)$, 
the left-handed positron~$E^c=(\tl{e}^c,e^c,F^{e^c})$, 
the goldstino supermultiplet~$Z=(z,\psi_z,F^z)$, 
the Higgs doublets~$H_i=(H_i,\tl{h}_i,F^{H_i})$ ($i=1,2$), and 
the gauge supermultiplets $V_1=(\lmd_1, A_{1\mu},D_1)$ and 
$V_2^a=(\lmd_2^a, A_{2\mu}^a,D_2^a)$ ($a=1,2,3$) for $U(1)_Y$ and $SU(2)_W$, 
respectively.  

The effective theory is described, up to higher-derivative terms, 
by a K\"{a}hler potential~$K$, a superpotential~$w$ 
and gauge kinetic functions~$f_i$ ($i=1,2$) as 
\be
 \cL = \int\! d^4\tht \;K +\sbk{\int\! d^2\tht \; w 
 +\int\! d^2\tht \; \frac{1}{4}\brc{f_1\cW_1^2+f_2\tr\brkt{\cW_2^2}}+\hc}, 
\ee
where $\cW_{1\alpha}$ and $\cW_{2\alpha}$ are superfield strength 
of $V_1$ and $V_2\equiv V_2^a\cdot(\sgm^a/2)$, respectively. 
Their general forms with the above field content are given by\footnote{
The most general form of the gauge kinetic function also contains 
the $SU(2)_L$-triplet term~\cite{brignole5}: 
$f^{{\rm (trp)}a}=(\gm_f^{\rm (trp)}/\Lmd^3)E^cL\sgm^aH_1+\cdots$. 
Contributions of such terms to EDM are similar to 
those of $\gm_{fi}$-terms in $f_i$.
Then, in order to simplify the discussion, 
we will neglect such terms in the following. 
} 
\bea
 K \eql |Z|^2+\bar{L}e^{2(-\frac{1}{2}g_1 V_1+g_2 V_2^a\cdot\frac{\sgm^a}{2})}L
 +\bar{E}^c e^{2g_1 V_1}E^c 
  -\frac{\alpha_z}{4\Lmd^2}|Z|^4 \nonumber\\
 &&-\frac{\alpha_l}{\Lmd^2}|Z|^2|L|^2-\frac{\alpha_{e^c}}{\Lmd^2}|Z|^2|E^c|^2 
  -\brkt{\frac{\gm_K}{2\Lmd^3}\bar{Z}^2LE^cH_1+\hc}+\cdots,  \nonumber\\
 w \eql -F^*Z-\frac{\sgm}{6}Z^3+y_eLE^cH_1-\frac{\rho}{\Lmd}ZLE^cH_1
  +\mu H_1H_2+\cdots, \nonumber\\
 f_i \eql 1+\frac{2\eta_i}{\Lmd}Z+\frac{\gm_{fi}}{\Lmd^3}LE^cH_1+\cdots, 
 \;\;\;(i=1,2)
 \label{def_fcns}
\eea
where $g_1$ and $g_2$ are the gauge coupling constants of $U(1)_Y$ and 
$SU(2)_L$ respectively, and $y_e$ is the Yukawa coupling constant 
for the electron. 
Complex parameters~$F$ and $\mu$ have a mass-dimension two and one 
respectively, and the other parameters are all dimensionless. 
$\Lmd$ denotes the cut-off scale of the effective theory and set to be real 
and positive. 
The ellipses denote terms that either do not play any role in the following 
discussion or can be eliminated by analytic field redefinitions. 

We will assume that the only Higgs fields have non-zero VEVs.  
\be
 \vev{H_1^0}=\frac{v_1}{\sqrt{2}}, \;\;\;
 \vev{H_2^0}=\frac{v_2}{\sqrt{2}}, 
\ee
The other fields do not have any non-zero VEVs \footnote{
As a result, the K\"{a}hler metric and the gauge kinetic functions are 
canonical at the vacuum, so that the component fields are all canonically 
normalized. 
}. 
Therefore, we can check that 
\be
 \vev{F^Z}=F. 
\ee
Namely, SUSY is broken spontaneously. 

Using the $SU(2)_L\times U(1)_Y$ gauge symmetry, we can set the Higgs 
VEVs~$v_1$, $v_2$ and the Yukawa coupling~$y_e$ to be real. 
Furthermore, we will set the parameter~$F$ to be real and positive 
by the field redefinition of $Z$. 
As a result, the complex parameters in the theory are $\gm_K$, 
$\sgm$, $\rho$, $\mu$, $\eta_i$, and $\gm_{fi}$. 

Next, let us see how the soft SUSY breaking parameters are expressed 
in terms of the above parameters. 
Considering the fact that SUSY is broken by the $F$-term of $Z$, 
the fermionic component~$\psi_z$ corresponds to the goldstino, 
and thus is massless\footnote{
Strictly speaking, $\psi_z$ is not the goldstino itself. 
Since $\vev{F^{H_1^0}}=-\mu^*v_2/\sqrt{2}$, $\vev{F^{H_2^0}}
=-\mu^*v_1/\sqrt{2}$, 
$\vev{D_1}=g_1(v_1^2-v_2^2)/4$ and $\vev{D_2^3}=-g_2(v_1^2-v_2^2)/4$ 
after the electroweak symmetry breaking, 
the genuine goldstino~$\tl{G}$ has its components also 
in the higgsinos and gauginos, and thus $\psi_z$ has a small mass term 
suppressed by $v_1v_2/\Lsb^2$ \cite{brignole5}. 
However, we will refer to $\psi_z$ as the `goldstino' in the following, 
because $\vev{F^Z}$ is the dominant source of SUSY breaking, 
\ie, $\Lsb\simeq \sqrt{F}$. 
}. 
Masses of the scalar partners of the goldstino, the {\it sgoldstino}, 
are given by 
\be
 m_S^2=\frac{\alpha_z}{\Lmd^2}F^2+|\sgm|F, \;\;\;
 m_P^2=\frac{\alpha_z}{\Lmd^2}F^2-|\sgm|F,  \label{sgold_mass}
\ee
where 
\be
 z\equiv\frac{e^{-i\vph/2}}{\sqrt{2}}(S+iP). \;\;\; 
 \brkt{\vph\equiv\arg(\sgm)}
\ee

The gaugino masses are 
\be
 M_i \equiv\frac{\eta_i}{\Lmd}F, 
\ee
and the $A$-parameter~$A_e$ is given by 
\be
 y_e A_e\equiv \frac{\rho}{\Lmd}F. 
\ee
The selectron mass matrix in the basis~$(\tl{e},\tl{e}^{c*})$ is 
\be
 \cM_{\tl{e}}^2=\mtrx{\mll^2+m_e^2}{\mrl^{2*}}{\mrl^2}{\mrr^2+m_e^2}, 
\ee
where $m_e\equiv y_e v_1/\sqrt{2}$ is the electron mass, and 
\bea
 \mll^2 \defa \frac{\alpha_l}{\Lmd^2}F^2, \;\;\;
 \mrr^2 \equiv \frac{\alpha_{e^c}}{\Lmd^2}F^2,  \nonumber\\ 
 \mrl^2 \defa m_e(A_e+\mu^*\tan\beta). \;\;\;
 \brkt{\tan\beta\equiv\frac{v_2}{v_1}}
\eea

Using these expressions, we can rewrite all the parameters 
except $\gm_K$ and $\gm_{fi}$ in Eq.(\ref{def_fcns}) 
in terms of the soft SUSY breaking parameters and 
the order parameter of SUSY breaking~$\sqrt{F}$. 

Then,the independent CP violating phases that are relevant to the following 
discussion are $\arg(M_i^*A_e)$, $\arg(M_i\mu)$, 
$\arg(M^2 e^{-i\vph})$, $\arg(M\gm_K)$ and $\arg(M^*\gm_{fi})$. 
The first two types of the phases are the usual MSSM ones, 
but the remaining ones are the new phases associated with 
the TeV-scale SUSY breaking.

\section{Electron EDM} \label{est_EDM}
Now we will estimate EDM of the electron. 
For simplicity, we will suppose that the gauge kinetic functions 
are common, and denote the common gaugino mass and the coupling constant as 
$M$ and $\gm_f$, respectively. 

For large $\tan\beta$, the conventional MSSM contribution to $d_e$ is 
given by~\cite{EDM}
\be
 \frac{d_e^{\rm (MSSM)}}{e}\simeq
 \frac{5\alpha\Im(M^*\mrl^2)}{96\pi \sin^2\tht_W |M|^4}, 
 \label{de_mssm}
\ee
where $\alpha$ is the fine structure constant and $\tht_W$ is 
the Weinberg angle. 
In this case, $\mrl^2\simeq m_e\mu^*\tan\beta$. 
Here, we have assumed that $|M|\simeq \mll\simeq \mrr\simeq |\mu|$, 
for simplicity. 
Since we are interested in the goldstino contributions besides 
the MSSM contribution, 
we will assume that the above $d_e^{\rm (MSSM)}$ is negligibly small, \ie, 
the relative phase between $M$ and $\mrl^2$ is tuned to be small, 
in the following. 

\begin{figure}
\leavevmode
\epsfysize=7cm
\centerline{\epsfbox{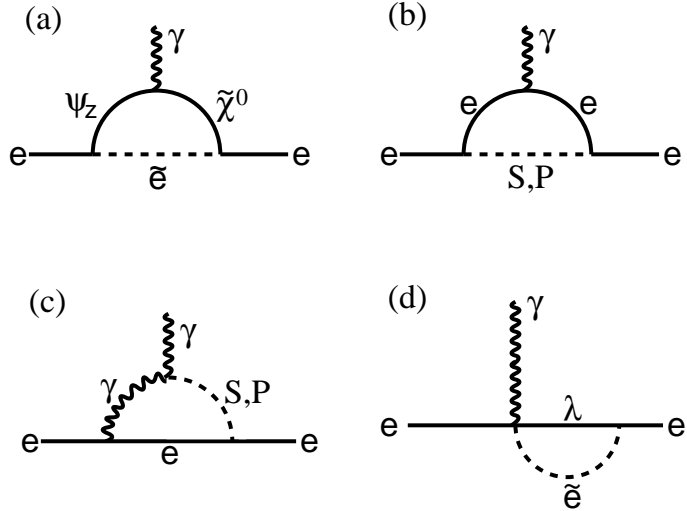}}
\caption{Additional diagrams to MSSM ones which contribute 
to the electron EDM. 
$\tl{\chi}^0$ stands for the neutralinos, and 
$\lmd$ is the photino. }
\label{rel_dgm_eEDM}
\end{figure}

Due to the interactions involving the goldstino supermultiplet~$Z$, 
there are additional contributions besides the usual MSSM ones. 
The relevant diagrams to $d_e$ are essentially the same as those to 
the anomalous magnetic moment~$a_\mu$, 
which are shown in Fig.~1 of Ref.\cite{brignole3}. 
Among them, four types of the diagrams shown in Fig.~\ref{rel_dgm_eEDM} 
can have a non-vanishing CP phases, and thus can contribute to EDM. 

Contributions of (a) and (b) to EDM are suppressed 
by $m_e^2$ compared to the other contributions, 
and thus can be neglected\footnote{
Here, we neglect regularization-dependent terms which emerge 
from the diagrams~(a) and (c) because they are cancelled 
with each other \cite{brignole3}. 
}. 
The remaining diagrams are logarithmically divergent. 
\be
 \frac{d_e^{(c)}}{e}=-\frac{m_e\Im(MA_e e^{-i\vph})}{16\pi^2 F^2}
 \ln\frac{m_S^2}{m_P^2}
 -\frac{\Im(\gm_K M)v_1}{16\sqrt{2}\pi^2\Lmd^3}
 \brkt{\ln\frac{\Luv^2}{m_S^2}+\ln\frac{\Luv^2}{m_P^2}}, 
 \label{de_c}
\ee
and 
\be
 \frac{d_e^{(d)}}{e}=-\frac{\Im(\gm_f M^*)v_1}{32\sqrt{2}\pi^2 \Lmd^3}
 \brc{\frac{|M|^2\ln(\Luv^2/|M|^2)-\mll^2\ln(\Luv^2/\mll^2)}{|M|^2-\mll^2} 
 +(\mll^2\exch\mrr^2)}, 
\ee
where $\Luv$ is a cut-off scale. 
Here, we have neglected terms suppressed by $m_e$, and assumed that 
the both sgoldstinos~$S$ and $P$ are much heavier than the electron. 

\begin{figure}
\leavevmode
\epsfysize=7cm
\centerline{\epsfbox{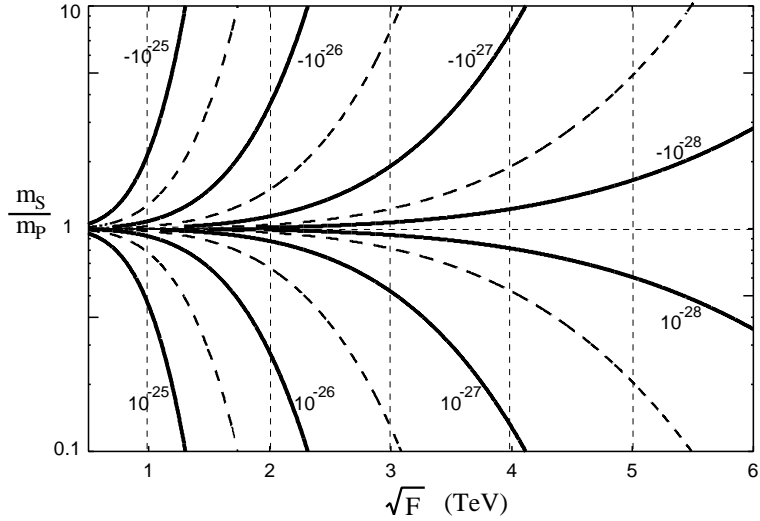}}
\caption{Constant contours of the value of $d_e$ on the 
$(\sqrt{F},m_S/m_P)$-plane.  The parameters are set as $|M|=A_e=1$~TeV, 
$\arg(MA_e e^{-i\vph})=\pi/2$, and $\gm_K=\gm_f=0$. 
The numbers in the plot shows the values of $d_e$ in units of $e$cm. }
\label{contour}
\end{figure}

First, we will consider the case that $\gm_K$ and $\gm_f$ are negligibly small. 
In this case, the result becomes finite and can be expressed 
by the SUSY breaking masses and the order parameter~$\sqrt{F}$. 
Fig.~\ref{contour} shows constant contours of the value of $d_e$ 
on the ($\sqrt{F}$, $m_S/m_P$)-plane. 
From this plot, we can see that the sgoldstino contribution can saturate 
the current experimental bound: $|d_e|<4.3\times 10^{-27}e$cm~\cite{commins}, 
when $\sqrt{F}$ is close to the soft mass scale. 

Notice that EDM vanishes when $m_S=m_P$. 
The reason for this is as follows. 
We have considered the case of $\gamma_K, \gamma_f \ll 1$, 
and thus the only CP phase 
besides the MSSM ones is $\vph$, the phase of 
the parameter~$\sgm$ in the superpotential. 
The degeneracy of the two sgoldstinos means that $\sgm=0$. 
(See Eq.(\ref{sgold_mass}). )
Therefore, when two sgoldstinos are degenerate, there exists no CP violating 
sources contributing to the electron EDM at one-loop level. 

So far, we have assumed that $\gm_K$ and $\gm_f$ are suppressed 
by some mechanism. 
If they are of order one, the resultant EDM value far exceeds its experimental 
upper bound. 
Then, to make the discussion complete, we will investigate the constraints 
on $\gm_K$ and $\gm_f$ from the experimental bound on $d_e$. 
Here, we will assume that all soft SUSY breaking masses are equal, 
for simplicity. 
Then, the expression of $d_e$ can be simplified as  
\be
 \frac{d_e}{e}\simeq -\frac{\Im(\gm_K M)v_1}{4\sqrt{2}\pi^2\Lmd^3}
 \ln\frac{\Lmd}{\tl{m}}
 -\frac{\Im(\gm_f^* M)v_1}{8\sqrt{2}\pi^2\Lmd^3}
 \brkt{\ln\frac{\Lmd}{\tl{m}}-\frac{1}{2}}, \label{de_gm}
\ee
where $\tl{m}\equiv m_S=m_P=|M|=\mll=\mrr$. 
Since the momentum cut-off scale~$\Luv$ is thought to be the same order 
as the scale parameter~$\Lmd$ that controls the higher-order 
terms in the effective Lagrangian, 
we have identified these two scales, \ie, $\Luv=\Lmd$. 

\begin{figure}
\leavevmode
\epsfysize=7cm
\centerline{\epsfbox{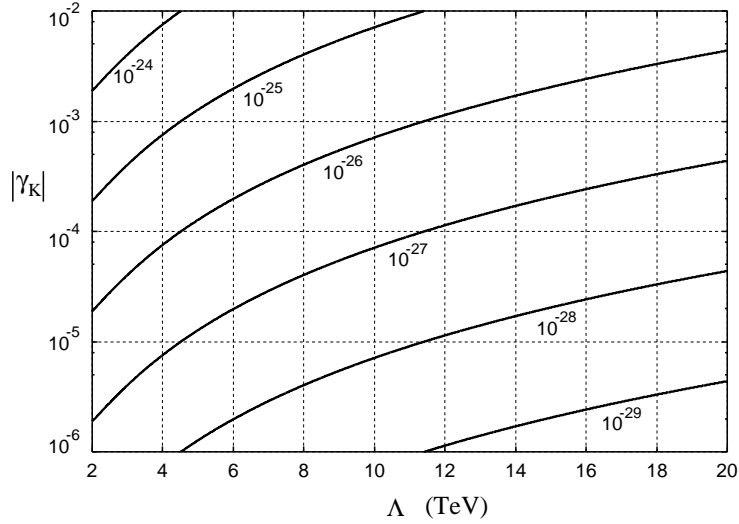}}
\caption{Constant contours of the value of $|d_e|$ on the 
$(\Lmd,|\gm_K|)$-plane. 
The parameters are set as $\tl{m}=1$~TeV, $\tan\beta=10$ and 
$\arg(\gm_K M)=\pi/4$. The numbers in the plot shows the values of 
$d_e$ in units of $e$cm. }
\label{contour2}
\end{figure}

Fig.~\ref{contour2} shows the value of $|d_e|$ on the ($\Lmd$,$|\gm_K|$)-plane 
calculated by Eq.(\ref{de_gm}) with $\gm_f=0$. 
We have set the parameters as $\tl{m}=1$~TeV and $\tan\beta=10$, 
and the CP phase is assumed to be unsuppressed, \ie, $\arg(\gm_K M)=\pi/4$. 
From the current experimental bound on $|d_e|$, 
the dimensionless parameter~$\gm_K$ must be suppressed at least 
by four to five orders of the magnitude when the cut-off scale~$\Lmd$ is 
in the range of 4-10 TeV, for example. 
For the smaller values of $\tan\beta$, the constraint on $\gm_K$ becomes 
severer. 
This constraint on $\gm_K$ can be realized if we introduce 
an approximate chiral symmetry that is broken explicitly by the order 
of the lepton masses \cite{brignole3}. 
In such a case, the flavor structure of $\gm_K$ is inferred as 
$m_\mu (\gm_K)_e \simeq m_e (\gm_K)_\mu$. 
Then, we can relate $d_e$ to the anomalous magnetic moment 
of the muon~$a_\mu$. 
That is, 
\be
 a_\mu^{\rm nonSM}\equiv a_\mu-a_\mu^{\rm SM}\simeq 
 \frac{2m_\mu^2}{m_e\tan\phi_K}\cdot\frac{d_e}{e}, 
\ee
where $a_\mu^{\rm SM}$ is the standard model contribution to $a_\mu$, 
$\phi_K\equiv\arg(\gm_K M)$ is the CP phase, which is set to be $\pi/4$ now. 
Using this relation, the contour of $10^{-27}e$cm in Fig.~\ref{contour2} 
corresponds to that of $2\times 10^{-12}$ for $a_\mu^{\rm nonSM}$. 
Therefore, we can see that the constraint on $\gm_K$ 
from the experimental bound for $a_\mu^{\rm nonSM}$ is weaker 
than that for $d_e$ by about three orders of magnitude\footnote{
We have used $\dlt a_\mu\equiv a_\mu^{\rm exp}-a_\mu^{\rm SM}<\cO(10^{-9})$ 
as a bound on $a_\mu^{\rm nonSM}$ \cite{bartos}. 
}. 

Similarly, the constraint on $\gm_f$ can be obtained by setting $\gm_K=0$ 
in Eq.(\ref{de_gm}). 
The required suppression of $\gm_f$ is roughly the same order as that of 
$\gm_K$.\footnote{Of course, these constraints become weaker 
if an accidental cancellation between two terms in Eq.(\ref{de_gm}) occurs.
} 

Although the naturalness considerations suggest that the sgoldstino masses 
are favored to be the same order as the slepton mass scale \cite{brignole6}, 
the possibility of the light sgoldstinos is not excluded if some dynamical 
mechanism that protects their small masses works. 
For instance, in the case that the sgoldstinos are interpreted as 
the (Pseudo-) Nambu-Goldstone 
bosons for some (approximate) global symmetries, 
their small masses are protected against the quantum corrections.  
In such a case, \ie, $m_S^2,m_P^2\ll m_e^2$, 
the sgoldstino contribution to EDM becomes 
\be
 \frac{d_e^{(c)}}{e}=-\frac{m_e\Im(M^* A_e)}{16\pi^2 F^2}. 
\ee
Note that this is proportional to the factor~$\Im(M^* A_e)$ that is common 
to the conventional MSSM contribution. 
So if there is some mechanism that suppresses the MSSM contributions, 
the above sgoldstino contribution also receives the same suppression. 

It is straightforward to extend the discussion to 
the neutron and the mercury EDMs. 
We can see that the situations become similar to 
that of the electron EDM.

\section{EDM in the warped Gaugino mediation scenario} \label{warped_GM}
In this section, we will consider the gaugino mediation model 
with the warped geometry \cite{GP} 
as an example of models with the TeV-scale SUSY breaking, 
and provide a rough estimation of the electron EDM. 

The background metric is 
\be
 ds^2=e^{-2k|y|}\eta_{\mu\nu}dx^\mu dx^\nu+dy^2, 
\ee
where $1/k$ is the AdS curvature radius, and $x^\mu$ ($\mu=0,1,2,3$) and $y$ 
are the coordinates of our four dimensions and the fifth dimension, 
respectively. 

The effective mass scale on the boundary at $y=0$ is the Planck mass~$\Mp$, 
while that on the boundary at $y=\pi R$ is $\Mp e^{-\pi kR}$, 
which will be associated with the TeV scale provided $kR\simeq 12$. 
So we will refer to the boundaries at $y=0$ and $y=\pi R$ as the Planck brane 
and the TeV brane, respectively. 
SUSY is supposed to be broken on the TeV brane, so that 
the order parameter of SUSY breaking~$\Lsb$ is $\cO({\rm TeV})$. 
We have assumed that the quark, the lepton and the Higgs supermultiplets 
are localized on the Planck brane and the gauge supermultiplets live 
in the bulk. 
Therefore, the SUSY breaking effects are mediated by the gauge interactions. 
Namely the gaugino masses are generated at tree-level, 
while the sfermion masses 
and the $A$-parameters are induced at one-loop level. 
As a result, the CP phases of the $A$-parameters are automatically 
aligned with those of the gauginos. 
So, the MSSM contributions to EDM are automatically suppressed in this model. 
On the other hand, as discussed in the previous section, 
the goldstino supermultiplet~$Z$ 
might provide a significant contributions to EDM 
because $\Lsb=\cO({\rm TeV})$ in this model. 
However, due to the separation along the fifth dimension, 
there are no direct couplings between $Z$ and the matter fields. 
So, possible one-loop diagrams are only those of type~(d) 
in Fig.\ref{rel_dgm_eEDM}. 
Contribution from them is, however, negligible 
due to the large suppression by $\Mp$ because the fundamental scale at $y=0$ 
is $\Mp$.
Therefore, there are no sizable contributions to EDM at one-loop level. 

\begin{figure}
\leavevmode
\epsfysize=6cm
\centerline{\epsfbox{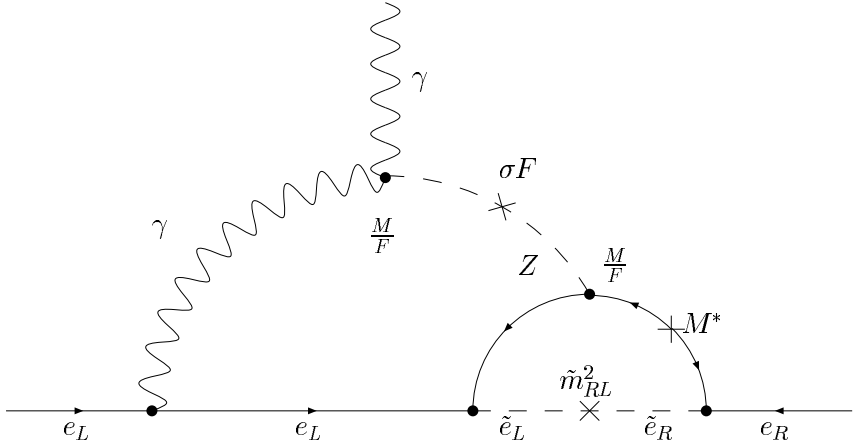}}
\caption{The diagram that provides the dominant contribution to $d_e$ 
in the large $\tan\beta$ case in the warped gaugino mediation scenario.  }
\label{2loop_dgm}
\end{figure}

The leading contributions to EDM come from the two-loop diagrams. 
For example, in the large $\tan\beta$ case, the diagram shown 
in Fig.~\ref{2loop_dgm} provides the dominant contribution to $d_e$. 

Since the gauge supermultiplets propagate in the bulk, there are an infinite 
number of the Kaluza-Klein (K.K.) modes in the four-dimensional perspective. 
However, such K.K. modes do not provide sizable contributions 
because their couplings to the matter fields, which are localized 
on the Planck brane are highly suppressed \cite{GP}. 
So, only the zero-modes contribute to the estimation of $d_e$. 

In this model, $\mrl^2\simeq m_e \mu^*\tan\beta$ 
since the $A$-parameter~$A_e$ is suppressed by the loop factor. 
Therefore, the electron EDM is roughly estimated as 
\be
 \frac{\abs{d_e}}{e}\simeq 
 \frac{\alpha}{(4\pi)^3\cos^2\tht_W F^2}
 \abs{\Im(M\mrl^2 e^{-i\vph})}
 \simeq \frac{\alpha m_e|\mu|\tan\beta}{(4\pi)^3\cos^2\tht_W |M|F^2}
 \abs{\Im(M^2e^{-i\vph})}. 
\ee
Here, we have assumed that $|M|^2\sim |\sgm F|=|m_S^2-m_P^2|/2$, 
and that the conventional MSSM CP phase~$\arg(M^*\mrl^2)$ is suppressed 
by some mechanism.   
For example, in the case that $|M|\sim |\mu|\sim \sqrt{F}\sim 1$~TeV 
and the CP phase is maximal, the estimated value of $d_e$ is 
\be
 |d_e|\sim 10^{-28}\tan\beta \;\mbox{($e$cm)}. 
\ee
Hence, the predicted value of $d_e$ can be close to its experimental bound 
in this model if the SUSY breaking scale~$\Lsb\simeq\sqrt{F}$ is close to 
the gaugino masses and $\tan \beta$ takes a value of $O(10)$.

\section{Summary and discussion} \label{conclusion}
In a class of models where SUSY is spontaneously broken at the TeV scale, 
interactions with the goldstino supermultiplet becomes 
relevant to low-energy observables. 
In this paper, we have investigated contributions 
of the goldstino supermultiplet to the electron EDM~$d_e$, 
which is one of the most sensitive observables for high-energy physics beyond 
the standard model. 
We found that the goldstino interactions induce sizable contributions to $d_e$, 
and they can saturate 
the experimental upper bound on $d_e$ 
even if the conventional MSSM contributions are suppressed 
when the SUSY breaking scale~$\Lsb$ is close to the soft mass scale. 

Concerning the relevant terms to $d_e$ at one-loop level, 
most of the parameters in the effective theory can be expressed in terms of 
the soft SUSY breaking parameters and $\Lsb$. 
Only two parameters~$\gm_K$ and $\gm_f$ are left as free parameters. 
However, these two parameters must be suppressed 
by the factor~$10^{-5}-10^{-4}$ 
due to the requirement that the predicted value of $d_e$ does not exceed 
the current experimental bound. 
Such suppression can be realized, for example, 
by introducing a chiral symmetry that is 
explicitly broken by the order of the lepton masses. 

One of the specific example of models with the TeV-scale SUSY breaking is 
the gaugino mediation scenario on the warped geometry. 
We have estimated the value of $d_e$ in such a model. 
Since there are no contributions to $d_e$ at one-loop level in this model, 
the dominant contributions are induced at two-loop level. 
Although they are suppressed by the loop factor, 
the value of $d_e$ can be close to the experimental bound 
especially in the large $\tan\beta$ case. 

Another possibility of the TeV-scale SUSY breaking is the existence of 
the strong coupling dynamics just above the TeV scale \cite{luty,NS}. 
Among such a class of models, a certain type of models 
can be interpreted as a dual theory of the above mentioned 
sequestered SUSY-breaking model with the warped geometry \cite{NS}, 
from the viewpoint of the AdS/CFT correspondence \cite{maldacena}. 
Furthermore, there is an interesting model that solves the strong CP problem 
in the context of the strong coupling dynamics \cite{hiller}. 
Applying our discussion to such a model is an intriguing subject. 

The search of flavor changing processes also provide valuable information 
on high-energy physics beyond the weak scale. 
In fact, the goldstino interactions contribute such processes, 
and their contributions can be sizable when $\Lsb$ is 
close to the soft mass scale \cite{brignole4}. 
In Ref.\cite{brignole4}, they focused the contributions to the flavor changing 
processes which depend on the interactions related to the mass spectra.  
In this case, the sources of flavor violation of the goldstino 
interactions are common to those of MSSM sector. 
Then, if there is some mechanism that suppresses the MSSM contributions 
to the flavor violation, the goldstino contributions also receive the same 
suppression. 
On the other hand, there is an additional CP source to the MSSM ones 
in the goldstino sector 
even in the absence of $\gm$-terms.  Thus, the goldstino contributions 
can saturate the experimental bound regardless of suppression mechanisms 
of the MSSM ones. 

The $\gm_K$ or $\gm_f$ term may also become the source of flavor violation. 
Then the goldstino contributions to the flavor violating processes become 
large as in the case of CP violation. However if the chiral suppression is 
assumed to suppress the contribution to the EDM, the constraints from 
the flavor changing processes are relaxed simultaneously.  
Moreover though the effective Lagrangian Eq. (\ref{def_fcns}) is the most 
general one in the context of the particle content given in this paper, 
there might be possible to introduce additional particles and interactions 
which contribute to the flavor changing processes.  
Those terms generally also become new sources of CP violation like 
the EDMs.  Thus CP violation is one of the most useful phenomena 
to detect a signal of the TeV-scale SUSY breaking. 

\vspace{5mm}

\begin{center}
{\bf Acknowledgments}
\end{center}
The authors thank Masahiro Yamaguchi for the useful discussion.  
M.E. is supported by the Japan Society for the Promotion of Science 
for Young Scientists.

\end{document}